\documentclass[10pt]{article}
\pdfoutput=1
\usepackage{graphics}
\usepackage[dvips]{graphicx}
\usepackage{ae}
\usepackage[T1]{fontenc}
\usepackage[ansinew]{inputenc}
\usepackage{amsmath}
\usepackage{amssymb}
\usepackage{graphicx}
\usepackage{caption}
\usepackage{color}
\usepackage[colorlinks]{hyperref}
\usepackage{lscape}
\usepackage{amsfonts}
\usepackage{amssymb}
\usepackage{makeidx}
\usepackage{braket}
\usepackage{bm}
\usepackage{bbm}
\usepackage{hyperref}
\usepackage{dsfont}
\usepackage{enumerate} 
\usepackage{float} 
\usepackage{mathrsfs}
\usepackage{authblk}
\usepackage{ulem}

\usepackage[a4paper,bindingoffset=0in,
            left=0.5in,right=0.5in,top=0.8in,bottom=0.8in,
            footskip=.25in]{geometry}
                                                                                                                                                                                                                              
\begin{document}

\title{Resonance interaction of two entangled atoms accelerating between two mirrors}
\author[1]{Riddhi Chatterjee \thanks{\emph{email:} riddhi.chatterjee@bose.res.in}}
\author[1]{Sunandan Gangopadhyay \thanks{\emph{email:} sunandan.gangopadhyay@gmail.com}}
\author[1]{A. S. Majumdar \thanks{\emph{email:} archan@bose.res.in}}
\affil{S. N. Bose National Centre for Basic Sciences, Block JD, Sector III, Salt Lake, Kolkata 700106, India.}                     
\date{}
%
\maketitle

\begin{abstract}
\small{We study the resonance interaction between two entangled identical atoms coupled to a quantized scalar field vacuum, and accelerating between two mirrors. We show how    radiative processes of the two-atom  entangled state can be manipulated by the atomic configuration undergoing noninertial motion. Incorporating the Heisenberg picture with symmetric operator ordering, the vacuum fluctuation and the self-reaction contributions are distinguished. We evaluate the resonance energy shift and the relaxation rate of energy of the two atom system from the self-reaction contribution in the Heisenberg equation of motion. We investigate the variation of these two quantities with relevant parameters such as acceleration, interatomic distance and position with respect to the boundaries. We show that both the energy level shift and the relaxation rate can be controlled by tuning the above parameters.} 
\end{abstract}
\newpage
\section{Introduction}
Atom-field interaction is a significant arena in the field of fundamental studies \cite{scu} as well as in quantum technology \cite{atomop}. Radiative atoms or quantum emitters are basic building blocks of quantum networks \cite{net}. Recent experimental developments in controlled atom-field interactions, such as for superconducting qubits \cite{sqbt}, and laser cooled ${}^{87}Rb$ atoms \cite{tel1}, provide us an excellent platform to perform quantum tasks \cite{optq} and for verification of fundamental phenomena \cite{fndm,svid}. Radiative properties of atoms play a key role in these contexts. Spontaneous emission affects the shape and temporal profile of the signal in matter-wave interferometry \cite{int}. The photons used to  encode and distribute information across quantum networks are created and controlled by atomic radiation \cite{fan,joh1}. 

The physics behind spontaneous emission of atoms is governed by the interplay of vacuum fluctuation \cite{vf1,vf2} and radiation reaction or self-reaction \cite{sr,min}.  An interesting phenomenon arises in case where more than one atom interacts with the quantized vacuum. The self-reaction contribution leads to exchange of real photons between two atoms (one in ground state and the other in excited state). This phenomenon is known as resonance interaction \cite{pas1,crg,salam,for,andr,agar} where the energy shift depends upon the interatomic distance. For separable states, resonance interaction is a fourth order effect in the coupling constant, and inversely proportional to square of the interatomic distance in the far zone limit \cite{berm,don,bcn,raf}. 

For correlated symmetric (superradiant) or antisymmetric (subradiant) states, resonance interaction is a second order effect in the coupling constant, and is inversely proportional to the
interatomic distance \cite{pas1,crg,salam,haakh1,haakh2,persi}. Studying the effects of initial atomic entanglement in context of resonance interaction is important because entanglement is a widely used resource in quantum information science. In this context we would like to emphasize that previous works have studied resonance interaction between two initially maximally entangled (symmetric or antisymmetric) atomic states. However, non-maximally entangled states are ubiquitous in practical information processing due to noise effects. 

The effects of acceleration on atom-field interactions is a fascinating topic in
its own right. The fundamental question of observer dependence of the particle content
of quantum vacuum in non-inertial settings has been much explored ever since the
pioneering works of Fulling \cite{fulling}, Davies \cite{davies} and Unruh \cite{unruh}.
However, observing vacuum fluctuations of the background thermal fields require
very high accelerations that may be difficult to achieve using current technology.
On the other hand, it has been observed recently that effects of acceleration on
resonance interactions could be more prominent even for much lower values of
acceleration \cite{pas2,pas3,zhou}, and hence, may be more amenable for experimental detection. Acceleration of atoms is a ubiquitous feature in various optical, magnetic and gravitational traps \cite{clock,bong,grav}. 

Quantum systems such as trapped ultracold atoms can be used to probe acceleration \cite{int,drag}. Technological development in the field of circuit quantum electrodynamics has allowed us to explore quantum systems undergoing relativistic motion that is difficult to achieve using mechanical mirrors. Relativistic acceleration in superconducting circuits is highly effective in generating quantum correlations \cite{alva}, performing quantum tasks \cite{tel2} and also simulating tests of quantum theory in gravitational backgrounds \cite{vcamp}. In such systems large acceleration can be simulated by ultrafast variation of qubit-field coupling and can reach upto $10^{17} m/s^2$ \cite{hac}. Embedded quantum simulator \cite{simul} can generate acceleration of much larger order than parametric coupling. Since radiative properties of atoms is an important area of interest and plays a significant role in the above mentioned phenomena, the study of resonance interactions in the relativistic non-inertial backgrounds seems quite feasible in practical set-ups.

The study of atom-photon interactions in confined or structured environments has
received focused attention in recent times. Due to innovation in nanofabrication
techniques \cite{vetsch,goban}, observation of atomic excitations in nanoscale
waveguides has been performed \cite{corzo}, using trapped atoms in optical nanofibres
\cite{thompson,solano}. A new field of waveguide quantum electrodynamics is emerging
through such studies, which has opened up the possibilities of exploration of
fundamental quantum optical aspects such as atom-photon lattices \cite{chang}, as
well as the potential of long-distance quantum communication enabled using atomic
ensembles \cite{duan,sang}. Boundary conditions play the principal role in realising relativistic quantum phenomena in superconducting circuits \cite{alva,tel2,vcamp,hac}. At the basic level, a waveguide or cavity imposes boundary conditions that affect radiative properties of atoms. The consideration of  boundary effects cannot be overemphasized while studying fundamental phenomena such as
resonant interactions of entangled atomic systems, or developing quantum protocols 
using such systems confined within cavities or waveguides.

The motivation for the present work is to investigate how radiative properties of 
entangled atoms change under the combined effect of boundary conditions and acceleration of atoms. Resonance interaction has been studied in case of static neutral atoms in structured environments \cite{band,cyl}. However, quantum mechanical particles such as atoms can never be absolutely static, but rather undergo accelerations in various confining potentials in realistic scenarios \cite{int,clock,bong,grav}. Radiative properties of atoms in relativistic background is also a significant area of interest in testing fundamental phenomena with ultracold atoms and superconducting circuits \cite{fndm,svid}. The energy level shift due to resonance interaction has been studied in the context of two correlated and accelerating atoms  in free space as well as in front of a single infinite mirror \cite{pas2,pas3}. The rate of change of energy due to resonance interaction of two entangled atoms accelerating in free space has also been studied recently \cite{zhou}. Employing the formalism proposed by Dalibard, Dupont-Roc and Cohen-Tannoudji (DDC) \cite{dal1,dal2}, it has been shown that vacuum fluctuation does not contribute to resonance interaction which is a nonthermal effect caused by the radiation field of an atom  perturbed by the radiation field of the second atom. However, the resonance interactions indeed depends upon the system acceleration \cite{pas2,pas3,zhou}. 

In this work we employ the DDC formalism to perform a detailed study of the resonance interaction of an entangled two-atom system which accelerates between two mirrors. A simple approximation for a waveguide or a cavity is two infinite parallel plates. Here we consider two entangled atoms accelerating between two reflecting infinite parallel plates. The direction of acceleration of the atoms is parallel to the mirror. The atomic system interacts with a quantized scalar vacuum in the presence of Dirichlet boundary conditions imposed by the two plates. For the purpose of the present study, the scalar field captures the essential physical features of electromagnetic interactions of neutral atoms, such as the distance dependent resonance energy, without having to take into
account the technical complications associated with the electromagnetic field.

Since boundary conditions modify the density of states of the radiation field, the response of the field depends upon the boundary parameters. There are two quantities of interest arising due to resonance interaction \cite{dal1,dal2}. The first is the modification of the energy spectrum of the system in terms of energy level shift. Secondly, the average rate of change of energy, or the relaxation rate of energy of the two atoms, affects atomic transition due to resonance interaction.  We calculate the resonance energy level shift and relaxation rate of energy of the two-atom system
which accelerates under the above boundary condition.

In the present analysis we are further motivated to study how the spatial
location of the atoms within the region confined by the two mirrors, and the
spatial configuration of the two-atom system  impact  their resonant interaction.
For this purpose we consider here two different configurations -- the line joining 
the two atoms is parallel  to the axis of the plates in one case, and perpendicular to the axis of the plates in the other case.   We explicitly evaluate the energy level shift of the atoms and their rate of change of energy due to resonance interaction with boundary parameters for both the configurations and study the variation of the above quantities with respect to interatomic separation, distance of the atoms from one of the plates, and separation between the plates. Within the DDC framework \cite{dal1,dal2}, our analysis further enables us to study as limiting cases 
the resonance dynamics in case of free space or a single plate \cite{pas2,pas3,zhou}.
Our results show how modification of acceleration and geometry of the atom-cavity system can enhance or diminish both resonance energy shift and relaxation rate of energy due to resonance interaction. 

The organization of the paper is as follows. In section II, we  present a 
brief overview of some essential features of the  DDC formalism  \cite{dal1,dal2,aud1,aud2} used to obtain general expressions for resonance energy level shift and relaxation rate of energy due to resonance interaction of a two-atom system.  In section III we first specify
the atomic configurations and spatial geometry used our subsequent analysis. We next     evaluate explicitly the resonance energy level shift and study its dependence on various parameters as mentioned above. 
Finally, we perform a similar
analysis of the relaxation rate of energy due to resonance interaction. We conclude
with a summary and outlook of our results in section IV.

\section{Heisenberg picture of two accelerating atoms interacting with quantized scalar field}

In order to understand spontaneous emission by atoms, the effects of vacuum fluctuation
\cite{vf1,vf2} and radiation reaction \cite{sr} have been historically invoked either separately, or together \cite{min}. The choice of ordering of the atomic and field operators in the Hamiltonian leads to an ambiguity between the contributions of vacuum fluctuation and
radiation reaction. In their seminal approach, DDC \cite{dal1,dal2} chose a particular
ordering which leads to Hermitian individual contributions in the Hamiltonian, and thereby, independent
physical attributes for both vacuum fluctuation and radiation reaction. This formulation has been successfully applied to distinguish the contribution of vacuum fluctuation and self reaction in radiative processes of one or more atoms \cite{pas1,jhe1,jhe2}.
In particular, energy shifts of atomic levels and radiative properties of atoms
undergoing noninertial motion have been described using the DDC formalism \cite{pas,zhuyulu,zhouyu}. Here we briefly outline this formalism in the context
of our analysis.


Let a system of two identical two level atoms (A and B) interact with a quantized massless real scalar field in its vacuum. Atoms are treated as pointlike systems with internal energy eigenstates $\lbrace \ket{g}, \ket{e} \rbrace$ and energy eigenvalues $\pm \dfrac{1}{2}\hbar \omega_0$. The atoms accelerate between two perfectly reflecting parallel plates placed at $z=0$ and $z=L$ and extends from $-\infty$ to $\infty$ along the x-y plane. The atoms move with constant interatomic separation along parallel trajectories $X_A(\tau)$ and $X_B(\tau)$ respectively, where $\tau$ is the proper time. It is to be noted that there would be transient effects when the acceleration of the atoms is suddenly turned on, and this could lead to interesting phenomena \cite{finite}. However, such effects die out in a very short time and the analysis of the present paper considers dynamics at much later times than the time scale of such transient effects. \par
The multipolar Hamiltonian describing the above phenomena, at a particular proper time slice $\tau$  \cite{zhou,aud1,aud2} (in natural units) in the co-moving frame of the two atoms is given by 
\begin{equation}
H(\tau) = \omega_{0} \sigma^{A}_{3}(\tau) + \omega_{0} \sigma^{B}_{3}(\tau) + \int d^3k \  \omega_{k} a^{\dagger}_{k} a_{k} \dfrac{dt}{d\tau} +  \lambda [\sigma^{A}_{2}(\tau) \phi(X_{A}(\tau)) + \sigma^{B}_{2}(\tau) \phi(X_{B}(\tau))]
\end{equation}
where $\phi(X)$ is the quantized scalar field with mode expansion \begin{equation}
\phi (X) = \int d^3k \dfrac{1}{\sqrt{2 (2\pi)^3 \omega_k}} [ a_{k}(t) e^{i \overrightarrow{k} \cdot \overrightarrow{x}} + a_{k}^{\dagger}(t) e^{- i \overrightarrow{k} \cdot \overrightarrow{x}}].
\end{equation}
$a_{k}(t)$, $a_{k}^{\dagger}(t)$ are creation and annihilation operators of the $k^{th}$ bosonic mode, $\lambda$ is the atom-field coupling constant, and $\sigma_2$ and $\sigma_3$ are the atomic pseudo spin operators  given by
\begin{eqnarray}
\sigma_3 &=& \dfrac{ 1}{2} (\ket{e}\bra{e}- \ket{g}\bra{g})\nonumber  \\
\sigma_2 &=&  \dfrac{i}{2}(\ket{g}\bra{e}- \ket{e}\bra{g}).
\end{eqnarray}
Applying the DDC formulation \cite{dal1,dal2}, the Heisenberg equation for reservoir (field) operator reads
\begin{align}\label{hr}
\dfrac{da_k(t(\tau))}{d\tau} &= i[H(\tau),a_{k}(t(\tau))] \nonumber \\
&= -i\omega_k a_{k}(t(\tau)) \dfrac{dt}{d\tau} + i\lambda \sigma_2^A [\phi(X_A(\tau)),a_{k}(t(\tau))] + 
i\lambda \sigma_2^B [\phi(X_B(\tau)),a_{k}(t(\tau))].
\end{align}
The Heisenberg equation for system (atom) operator reads \begin{eqnarray}
\dfrac{d\sigma_3(\tau)}{d\tau} = i[H(\tau),\sigma_3(\tau)]
&=& i\lambda [\sigma_2(\tau) , \sigma_3(\tau)] \phi(X(\tau)) \label{hs1} \\
\dfrac{d\sigma_{\pm}(\tau)}{d\tau} = i[H(\tau),\sigma_{\pm}(\tau)]
&=& i\omega_0 [\sigma_3(\tau) , \sigma_{\pm}(\tau)] + i\lambda [\sigma_2(\tau) , \sigma_{\pm}(\tau)] \phi(X(\tau)).\qquad \label{hs2}
\end{eqnarray}
where $\sigma_+ =\ket{e}\bra{g} $ and $\sigma_- = \ket{g}\bra{e}$.
Solution to equations (\ref{hr}),(\ref{hs1}),(\ref{hs2}) can be obtained by integrating them. The solutions (upto first order in $\lambda$) with distinct free (f) and source (s) part are given by \begin{eqnarray}
a_{k} (t(\tau)) &=&  a_{k}^{f} (t(\tau)) + a_{k}^{s} (t(\tau)) \nonumber \\
a_{k}^{f} (t(\tau)) &=& a_{k} (t(\tau_{0})) e^{- i \omega_{k} (t(\tau) - t(\tau_{0}))} \nonumber \\
a_{k}^{s} (t(\tau)) &=& i \lambda \lbrace \int^{\tau}_{\tau_{0}} d\tau' \sigma_2^A [\phi(X_{A}(\tau')), a_{k}^f (t(\tau))] + A\leftrightarrows B \  \text{term} \rbrace.
\end{eqnarray}
\begin{eqnarray}
\phi(X(\tau)) &=&  \phi^{f} (X(\tau)) + \phi^{s} (X(\tau)) \nonumber \\
\phi^f (X) &=& \int d^3k \dfrac{1}{\sqrt{2\omega_k}} [ a_{k}(t) e^{i \overrightarrow{k} \cdot \overrightarrow{x}} + a_{k}^{\dagger}(t) e^{- i \overrightarrow{k} \cdot \overrightarrow{x}}] \nonumber \\
\phi^{s} (X(\tau))&=& i \lambda \lbrace \int^{\tau}_{\tau_{0}} d\tau' \sigma_2^A [\phi(X_{A}(\tau')), \phi^{f} (X(\tau))] + A\leftrightarrows B \  \text{term} \rbrace.
\end{eqnarray}
\begin{eqnarray}
\sigma_3 (t(\tau)) &=&  \sigma_3^{f} (t(\tau)) + \sigma_3^{s} (t(\tau)) \nonumber \\
\sigma_3^{f} (t(\tau)) &=& \sigma_3^f (t(\tau_{0}))\nonumber  \\
\sigma_3^{s} (t(\tau)) &=& - i \lambda \int^{\tau}_{\tau_{0}} d\tau' \phi(x(\tau')) [\sigma_2^{f} (t(\tau')), \sigma_3^{f} (t(\tau))]. 
\end{eqnarray}
\begin{eqnarray}
\sigma_{\pm} (t(\tau)) &=&  \sigma_{\pm}^{f} (t(\tau)) + \sigma_{\pm}^{s} (t(\tau))  \nonumber \\
\sigma_{\pm}^{f} (t(\tau)) &=& \sigma_{\pm}^f (t(\tau_{0})) e^{- i \omega_{k} (t(\tau) - t(\tau_{0}))} \nonumber \\
\sigma_{\pm}^{s} (t(\tau)) &=& - i \lambda \int^{\tau}_{\tau_{0}} d\tau' \phi(x(\tau')) [\sigma_{2}^{f} (t(\tau')), \sigma_{\pm}^{f} (t(\tau))]. 
\end{eqnarray}
Let us consider the Heisenberg equation of motion of an arbitrary system observable $G_A(\tau)$ of system A (in principle it can be any system observable of either of the systems). Choosing symmetric operator ordering between atom and field variables and using the above solutions, the rate of change of G (upto second order in $\lambda$), averaged over the reservoir variable is given by 
\begin{equation}\label{rate}
\begin{split}
\Big\langle \Big( \dfrac{dG}{d\tau} \Big)_{vf} \Big \rangle_{\phi} = - \lambda^2 \int_{\tau_0}^{\tau} d\tau' C^F(X_A(\tau),X_A(\tau')) [\sigma_2^{A,f} (t(\tau')),[\sigma_2^{A,f} (t(\tau)),G_A^f(\tau)]] \\[10pt]
\Big\langle \Big( \dfrac{dG}{d\tau} \Big)_{sr} \Big \rangle_{\phi} = - \lambda^2 \int_{\tau_0}^{\tau} d\tau' [\chi^F(X_A(\tau),X_A(\tau')) [\sigma_2^{A,f} (t(\tau')),[\sigma_2^{A,f} (t(\tau)),G_A^f(\tau)]] + \\[3pt]
\chi^F(X_B(\tau),X_A(\tau')) [\sigma_2^{B,f} (t(\tau')),[\sigma_2^{A,f} (t(\tau)),G_A^f(\tau)]]].
\end{split}
\end{equation}
where `vf' and `sr' imply vacuum fluctuation and self reaction, respectively. $C^{F}$ and $\chi^{F}$ are given by
\begin{equation}
C^{F}(X(\tau),X(\tau')) = \dfrac{1}{2} \braket{0\mid \lbrace \phi^{F}(X(\tau)), \phi^{F}(X(\tau')) \rbrace \mid 0}
\end{equation}
and
\begin{equation}
\label{asym}
\chi^{F}(X(\tau),X(\tau')) = \dfrac{1}{2} \braket{0\mid [ \phi^{F}(X(\tau)), \phi^{F}(X(\tau')) ] \mid 0}
\end{equation}
where $\ket{0}$ denotes the Minkowski vacuum. $C^F$ is the symmetric field correlation and $\chi^F$ is the anti-symmetric field correlation or susceptibility. 
The commutator part of  Eq.(\ref{rate}) is known as the effective Hamiltonian  given by
 \begin{eqnarray}
(H_{A,eff}) = (H_{A,eff})_{vf} + (H_{A,eff})_{sr} \nonumber \\
(H_{A,eff})_{vf} = - \dfrac{i \lambda^2}{2} \int^{\tau}_{\tau_{0}} d\tau' C^{F}(X_{A}(\tau),X_{A}(\tau'))  \left[\sigma_{2 }^{A,f} (t(\tau)), \sigma_{2}^{A,f} (t(\tau'))\right]
 \\[6pt]
(H_{A,eff})_{sr} = - \dfrac{i \lambda^2}{2} \int^{\tau}_{\tau_{0}} d\tau' \Big[\chi^{F}(X_{A}(\tau),X_{A}(\tau')) \left[\sigma_{2}^{A,f} (t(\tau)), \sigma_{2}^{A,f} (t(\tau'))\right]\nonumber  \\ + \chi^{F}(X_{A}(\tau),X_{B}(\tau')) \left[\sigma_{2}^{A,f} (t(\tau)), \sigma_{2}^{B,f} (t(\tau'))\right]\Big] \nonumber
\end{eqnarray}

\subsection{Energy level shift}
The shift in energy due to the effective Hamiltonian can be calculated using non-degenerate perturbation theory \cite{pas2}. The total energy level shift of the two particle system is the sum of energy shifts of each system \begin{align} \label{rshift}
\delta E &= (\delta E)_{vf} + (\delta E)_{sr}  \nonumber \\
(\delta E)_{vf} &= -i\lambda^2 \int^{\tau}_{\tau_{0}} d\tau' [C^{F}(X_{A}(\tau),X_{A}(\tau')) \chi_A(\tau,\tau') + A\leftrightarrows B \  \text{term}]  \nonumber \\
(\delta E)_{sr} &= -i\lambda^2 \int^{\tau}_{\tau_{0}} d\tau' [\chi^{F}(X_{A}(\tau),X_{A}(\tau')) C_A(\tau,\tau')  \nonumber \\ &+ \chi^{F}(X_{A}(\tau),X_{B}(\tau')) C_{A,B}(\tau,\tau') + A\leftrightarrows B \  \text{term}] 
\end{align}
where $C(\tau, \tau')$ and $\chi(\tau, \tau')$ are the symmetric and anti-symmetric atomic correlations, given by
\begin{equation}
\chi(\tau, \tau') = \dfrac{1}{2} \braket{\psi \mid [\sigma_{2}^{f} (t(\tau)), \sigma_{2}^{f} (t(\tau'))] \mid \psi}
\end{equation}
\begin{equation}\label{sym}
C(\tau, \tau') = \dfrac{1}{2} \braket{\psi \mid \lbrace \sigma_{2}^{f} (t(\tau)), \sigma_{2}^{f} (t(\tau'))\rbrace \mid \psi}.
\end{equation}
$(\delta E)_{vf}$ and the first term of $(\delta E)_{sr}$ (including their $A\leftrightarrows B$ counterparts) represent Lamb shifts. The second term of $(\delta E)_{sr}$ (and their $A\leftrightarrows B$ counterparts) represent resonance energy shift. $\ket{\psi}$ denotes the state of the two particle system.

\subsection{Rate of change of energy}
Substituting $G(\tau)$ in Eq.(\ref{rate}) by the system Hamiltonian $H_s(\tau) = \omega_{0} \sigma^{A}_{3}(\tau) + \omega_{0} \sigma^{B}_{3}(\tau) $ and taking average with respect to the state $\ket{\psi}$, we get the total rate of change of energy of the system to be \begin{align} \label{exch}
R &= R_{vf}  + R_{sr} \nonumber \\
R_{vf} &=  2i\lambda^2 \int^{\tau}_{\tau_{0}} d\tau'  [C^{F}(X_{A}(\tau),X_{A}(\tau'))  \dfrac{d}{d\tau} \chi_A(\tau,\tau')  + A\leftrightarrows B \  \text{term}] \nonumber \\
R_{sr} &=  2i\lambda^2 \int^{\tau}_{\tau_{0}} d\tau' [\chi^{F}(X_{A}(\tau),X_{A}(\tau')) \dfrac{d}{d\tau} C_A(\tau,\tau') \nonumber \\ &+ \chi^{F}(X_{A}(\tau),X_{B}(\tau')) \dfrac{d}{d\tau} C_{A,B}(\tau,\tau') + A\leftrightarrows B \  \text{term}] 
\end{align}
where $R = \Big\langle \Big( \dfrac{dH_s(\tau)}{d\tau} \Big) \Big \rangle_{\phi,\psi}$, $R_{vf} = \Big\langle \Big( \dfrac{dH_s(\tau)}{d\tau} \Big)_{vf} \Big \rangle_{\phi,\psi}$, $R_{sr} = \Big\langle \Big( \dfrac{dH_s(\tau)}{d\tau} \Big)_{sr} \Big \rangle_{\phi,\psi}$. $R_{vf}$ and the first term of $R_{sr}$ (with their $A\leftrightarrows B$ counterparts) represent the rate of change of energy due to spontaneous emission. The second term of $R_{sr}$ (with their $A\leftrightarrows B$ counterparts) represents the rate of change of energy due to resonance interaction.

\section{Resonance energy shift and relaxation rate of two entangled atoms accelerating between parallel mirrors}

\begin{figure}[!htb]
\centering
\includegraphics[scale=0.3]{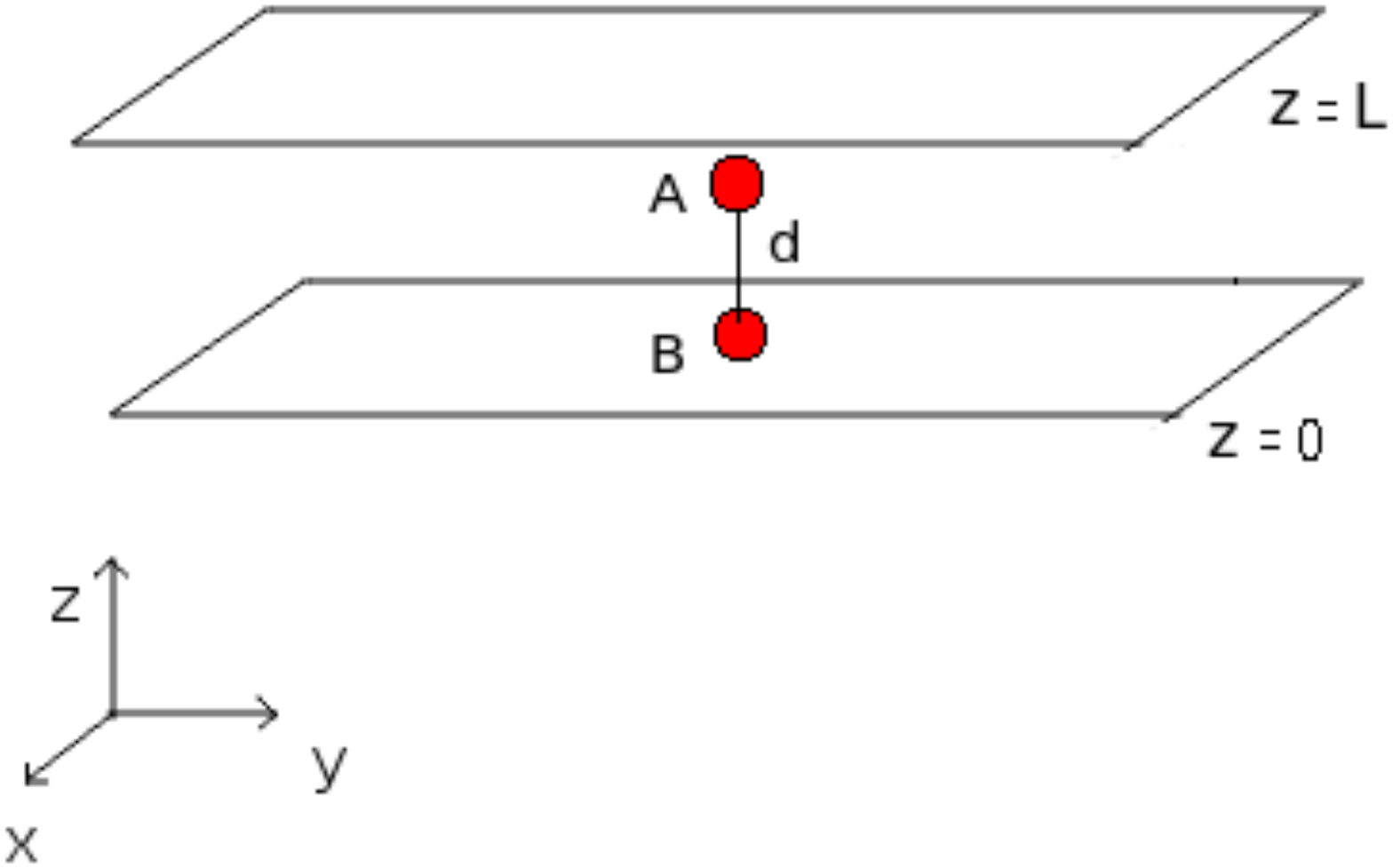}
\caption{(Color Online) Line joining two atoms perpendicular to the plates}
\label{pictest1}
\end{figure}

\begin{figure}[!htb]
\centering
\includegraphics[scale=0.3]{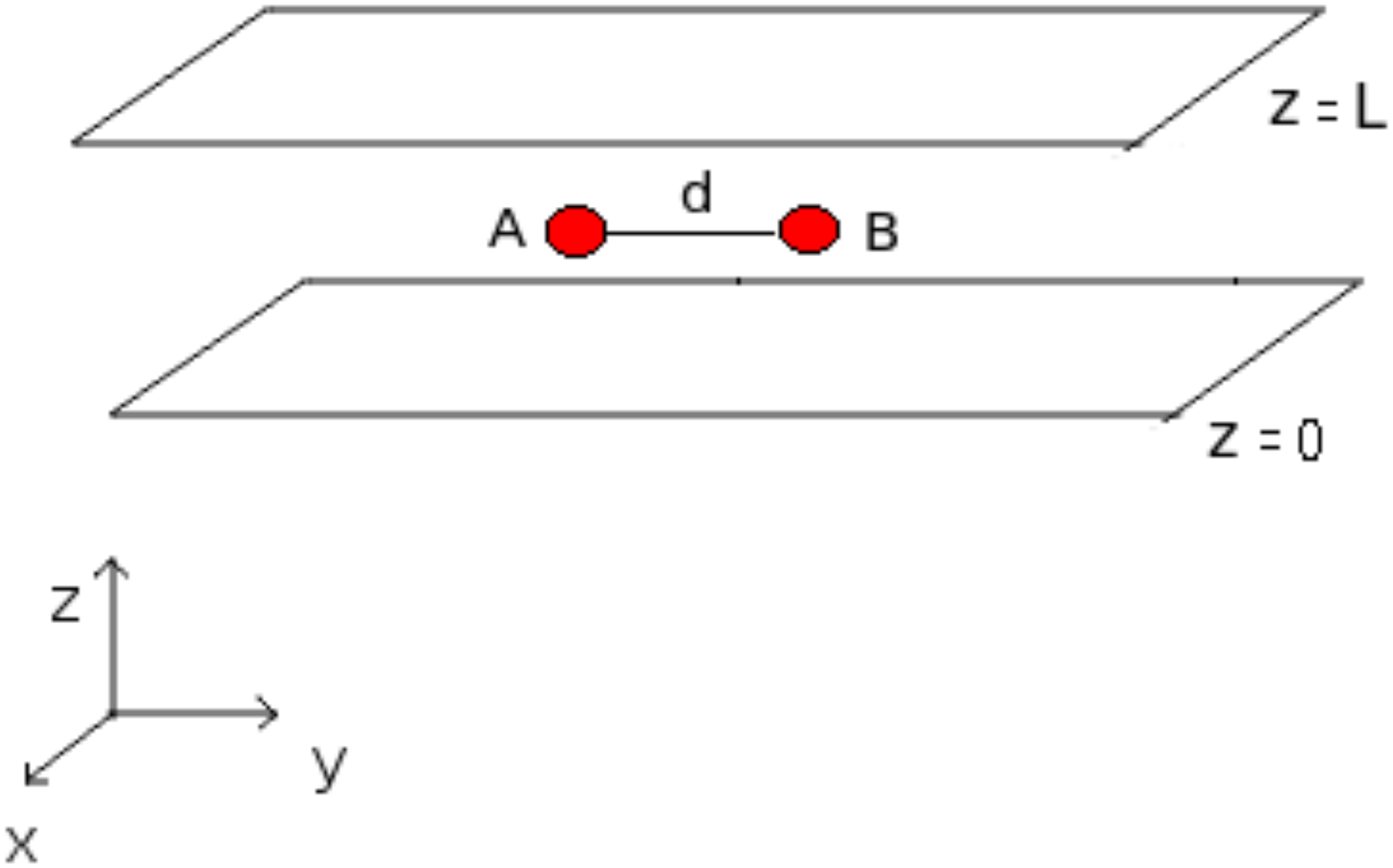}
\caption{(Color Online) Line joining two atoms parallel to the plates}
\label{pictest2}
\end{figure}


Let us now specify the  details of our set-up.
We choose the quantum state of the two atoms to be \begin{equation}\label{atst}
\ket{\psi} = \sin \theta \ket{g_A,e_B} + \cos \theta \ket{e_A,g_B}
\end{equation}
where $0 \leq \theta \leq \pi$. So, $\theta = 0, \pi/2, \pi $ represent separable states, and
the maximally entangled states are given by $\theta = \dfrac{\pi}{4}$ (superradiant state), and $\theta = \dfrac{3\pi}{4}$ (subradiant state). 

The plates are located at $z=0$ and $z=L$, respectively. We consider the following two different configurations. 
\paragraph*{\textbf{Configuration 1:}} The line joining the two atoms is perpendicular to the plates (Figure \ref{pictest1}). Trajectories $(X_{A/B})$ of the particles are given by \begin{equation}\label{con1}
\begin{split}
t_{A/B}(\tau) = \dfrac{1}{a} \sinh{a \tau} \qquad  x_{A/B}(\tau) = \dfrac{1}{a} \cosh{a \tau} \\
y_A = y_B = 0 \qquad z_A = z_0, z_B = z_0 +d
\end{split}
\end{equation}
where $a$ is the proper acceleration of both the atoms, and $d$ is the interatomic distance.

\paragraph*{\textbf{Configuration 2:}} The line joining the two atoms is parallel to the plates (Figure \ref{pictest2}). Trajectories $(X_{A/B})$ of the particles are given by
 \begin{equation}\label{con2}
\begin{split}
t_{A/B}(\tau) = \dfrac{1}{a} \sinh{a \tau} \qquad  x_{A/B}(\tau) = \dfrac{1}{a} \cosh{a \tau} \\
y_B = y_A + d \qquad  z_A = z_B = z_0.
\end{split}
\end{equation} 

Our goal is to evaluate the energy level shift and relaxation rate of energy due to resonance interaction. In order to do so we need to calculate $\chi^{F}(X_{A}(\tau),X_{B}(\tau'))$ and $C_{A,B}(\tau,\tau')$.
 Recalling eq. (\ref{asym}) we write, \begin{equation}
\chi^{F}(X_A(\tau),X_B(\tau')) = \dfrac{1}{2} \braket{0\mid [ \phi^{F}(X_A(\tau)), \phi^{F}(X_B(\tau')) ] \mid 0}. 
 \end{equation}  
The Wightman function between two space-time points in presence of two mirrors is 
derived using the method of images and is the result of multiple reflections of the photons between the two mirrors. It is
given by \cite{riz,dav,brw}
\begin{eqnarray}
W(X(\tau),X(\tau')) &=& \braket{0\mid [ \phi^{F}(X(\tau)) \phi^{F}(X(\tau')) ] \mid 0} \nonumber \\
&=& -\dfrac{1}{4\pi^2} \sum_{n=-\infty}^{\infty} \Big[ \dfrac{1}{(\Delta t - i\eta)^2 -\Delta x^2 - \Delta y^2 - (2 L n - \Delta z)^2} \nonumber \\
&-& \dfrac{1}{(\Delta t - i\eta)^2 -\Delta x^2 - \Delta y^2 - (2 L n - z - z')^2} \Big].
\end{eqnarray}

Using above correlation the susceptibility can be written as 
\begin{equation} \label{sus}
\begin{split}
\chi^{F}(X(\tau),X(\tau')) = -\dfrac{i}{4\pi} \sum_{n=-\infty}^{\infty} \epsilon (\Delta t) \Big[ \delta(\Delta t^2 -\Delta x^2 - \Delta y^2 - (2 L n - \Delta z)^2)  \\ -  \delta(\Delta t^2 -\Delta x^2 - \Delta y^2 - (2 L n - z - z')^2) \Big]
\end{split}
\end{equation}
where \begin{equation}
\begin{split}
\epsilon(\Delta t) = 1 \quad\ \  \text{for} \ \  \Delta t > 0 \\
= -1 \quad \text{for} \  \Delta t < 0.
\end{split}
\end{equation}
Substituting the particle trajectories from Eqs. (\ref{con1},\ref{con2}) in Eq. (\ref{sus}) and using the relation $\delta(f(r)) = \sum_j \dfrac{\delta(r - r_j)}{ \mid f'(r_j) \mid} $ (where $r_j$(s) are the roots of $f(r)$), we evaluate the susceptibility
for the respective configurations. 
\paragraph*{\textbf{Configuration 1:}} 
\begin{align}\label{chno}
\chi^{F}_{\perp}(X_A(\tau),X_B(\tau')) &= 
- \dfrac{1}{8\pi^2} \int_0^{\infty} d\omega (e^{i\omega \Delta \tau} - e^{-i\omega \Delta \tau}) \times  \nonumber \\
&\sum_{n=-\infty}^{\infty} \Big[\dfrac{\sin (\dfrac{2\omega}{a}\sinh^{-1}(\dfrac{z_1 a}{2}))}{z_1 \sqrt{1+\dfrac{z_1^2 a^2}{4}}}  - \dfrac{\sin (\dfrac{2\omega}{a}\sinh^{-1}(\dfrac{z_2 a}{2}))}{z_2 \sqrt{1+\dfrac{z_2^2 a^2}{4}}} \Big]
\end{align}
where $\Delta\tau = (\tau - \tau')$, $z_1 = \mid 2nL - d \mid$ and $z_2 = \mid 2nL -2z_0 -d \mid$.
\paragraph*{\textbf{Configuration 2:}} 
\begin{align}
\label{chpa}
\chi^{F}_{\parallel}(X_A(\tau),X_B(\tau')) &=  - \dfrac{1}{8\pi^2} \int_0^{\infty} d\omega (e^{i\omega \Delta \tau} - e^{-i\omega \Delta \tau}) \times  \nonumber \\
&\sum_{n=-\infty}^{\infty} \Big[\dfrac{\sin (\dfrac{2\omega}{a}\sinh^{-1}(\dfrac{z_3 a}{2}))}{z_3 \sqrt{1+\dfrac{z_3^2 a^2}{4}}} - \dfrac{\sin (\dfrac{2\omega}{a}\sinh^{-1}(\dfrac{z_4 a}{2}))}{z_4 \sqrt{1+\dfrac{z_4^2 a^2}{4}}} \Big]
\end{align}
where $z_3 = \sqrt{d^2 + 4n^2L^2}$ and $z_4 = \sqrt{d^2 + 4(nL-z_0)^2}$.

In order to obtain the atomic correlation,
using Eq.(\ref{sym}) we first write
\begin{equation}\label{atco}
C_{A,B}(\tau, \tau') = \dfrac{1}{2} \braket{\psi \mid \lbrace \sigma_{2}^{A,f} (t(\tau)), \sigma_{2}^{B,f} (t(\tau'))\rbrace \mid \psi}.
\end{equation}
Next, from Eqs.(\ref{atco}, \ref{atst}) we get \begin{equation}\label{cab}
C_{A,B} (\tau,\tau') = \dfrac{\sin 2\theta}{8} (e^{i\omega_0 \Delta \tau} - e^{-i\omega_0 \Delta \tau}).
\end{equation}
 
\subsection{Resonance energy shift}
Applying Eq. (\ref{rshift}) for the case of the two particles A and B, we obtain the resonance energy shift $\delta E_r$ for the two atom system, given by  \begin{equation}
\begin{split}
\delta E_r = -i\lambda^2 \int^{\tau}_{\tau_{0}} d\tau' [\chi^{F}(X_{A}(\tau),X_{B}(\tau')) C_{A,B}(\tau,\tau') + A\leftrightarrows B \  \text{term}]. 
\end{split}
\end{equation}
Following the DDC formalism, the limits of the integral can be set as $\tau \rightarrow \infty$ and $\tau_0 \rightarrow - \infty$,  signifying that the time of flight of the atoms $(\tau - \tau_0)$ is much larger than the correlation time between the two atoms,
which gives a non-vanishing contribution to the above integral. However, it should also be kept in mind that $(\tau - \tau_0)$ is much smaller compared to the relaxation time of the atoms \cite{dal2}.  Using Eqs. (\ref{chno},\ref{chpa},\ref{cab}), we obtain the resonance energy shifts for the two respective configurations.
\paragraph*{\textbf{Configuration 1:}} 
\begin{equation}\label{shftn}
\begin{split}
(\delta E_r)_{\perp} = - \dfrac{\lambda^2 \sin 2\theta}{16 \pi} \sum_{n=-\infty}^{\infty} \Big[\dfrac{\cos (\dfrac{2\omega_0}{a}\sinh^{-1}(\dfrac{z_1 a}{2}))}{z_1 \sqrt{1+\dfrac{z_1^2 a^2}{4}}}  -  \dfrac{\cos (\dfrac{2\omega_0}{a}\sinh^{-1}(\dfrac{z_2 a}{2}))}{z_2 \sqrt{1+\dfrac{z_2^2 a^2}{4}}} \Big]
\end{split}
\end{equation}
\paragraph*{\textbf{Configuration 2:}} \begin{equation}\label{shftp}
\begin{split}
(\delta E_r)_{\parallel} = - \dfrac{\lambda^2 \sin 2\theta}{16 \pi} \sum_{n=-\infty}^{\infty} \Big[\dfrac{\cos (\dfrac{2\omega_0}{a}\sinh^{-1}(\dfrac{z_3 a}{2}))}{z_3 \sqrt{1+\dfrac{z_3^2 a^2}{4}}}  -  \dfrac{\cos (\dfrac{2\omega_0}{a}\sinh^{-1}(\dfrac{z_4 a}{2}))}{z_4 \sqrt{1+\dfrac{z_4^2 a^2}{4}}} \Big].
\end{split}
\end{equation}

\par Let us now calculate low acceleration limit of resonance energy shift. When acceleration is low  $a \ll L^{-1}, d^{-1}, z_0^{-1}$ i.e, $aL$, $az_0$, $ad \ll 1$, we find  \begin{align}\label{lim}
&((\delta E_r)_{\perp})_{\text{low acc.}} = \frac{\lambda^2 \sin{2\theta}}{16\pi} \sum_{i=1,2} (-1)^i \Bigg[ \sum_{n=1}^{\infty}\Big[ \frac{\cos{(\omega_0 M_i)}}{M_i} -\frac{a^2}{8} \Big( M_i \cos{(\omega_0 M_i)}  \nonumber \\ &- \dfrac{M_i^2 \omega_0}{3} \sin{(\omega_0 M_i)} \Big) \Big] + \sum_{n=0}^{\infty} \Big[ \frac{\cos{(\omega_0 Q_i)}}{Q_i} -\frac{a^2}{8} \Big( Q_i \cos{(\omega_0 Q_i)}  - \dfrac{Q_i^2 \omega_0}{3} \sin{(\omega_0 Q_i)} \Big) \Big] \Bigg]
\end{align}
where $M_1 = (2nL-d)$, $M_2 = (2nL-2z_0-d)$, $Q_1 = (2nL+d)$ and $Q_2 = (2nL+2z_0 +d)$, and we have retained terms upto order $a^2$. Low acceleration limit for the other configurations can be calculated in similar manner. Note that no term of linear order in $a$ is present. This resembles a previous result of inertial limit \cite{cai}.

We study the variation of  energy shift  due to resonance interaction for both configurations  with initial acceleration of the atoms ($a$), interatomic distance ($d$), and distance of the atoms from the mirror ($z_0$). 
The results are plotted below, where all physical quantities are expressed in dimensionless units. We choose a similar  order of magnitude for $\omega_0 L$, $\omega_0 z_0$ and $\omega_0 d$, since cavity effects are relevant when these length scales are comparable \cite{donaire}.

\begin{figure}[!htb]
\centering
\includegraphics[scale=0.4]{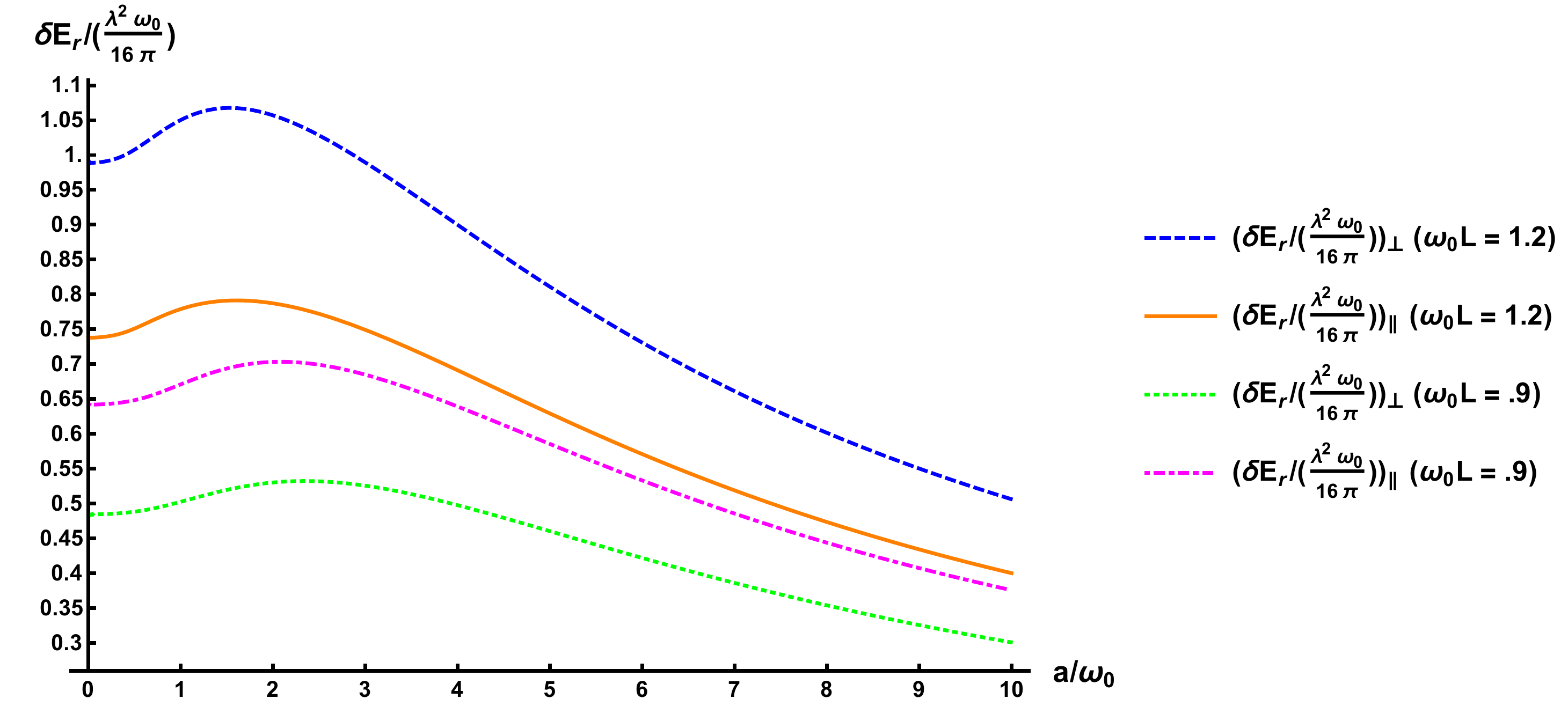}
\caption{(Color Online) Resonance energy level shift(per unit $\left(\frac{\lambda^2 \omega_0}{16 \pi}\right)$) versus acceleration ($(E_r/\left(\frac{\lambda^2 \omega_0}{16 \pi}\right))_{\perp}$ stands for atoms aligned perpendicular to the plates, $(E_r/\left(\frac{\lambda^2 \omega_0}{16 \pi}\right))_{\parallel}$ stands for atoms aligned parallel to the plates), $\theta = 3 \pi / 4$, $\omega_0 d = 0.5$, $\omega_0 z_0 = 0.3$.}
\label{eva}
\end{figure}

The resonance interaction depends upon the correlation between two atoms. In reality entanglement is not always preserved during experimental situations, and in many cases nonmaximally entangled states are used as probe or resource. Equations (\ref{shftn},\ref{shftp}) show that resonance energy shift varies sinusoidally with entanglement. In the figures presented here we plot the energy level shifts for the maximum values of interatomic entanglement.

First we study the variation of energy level shift due to varying acceleration. As mentioned earlier, atoms may be subjected to acceleration inside waveguides by the background potential. We consider here the case of maximally entangled atoms.
Since the magnitude of energy shift is same at $\theta =\frac{\pi}{4}$ and $\frac{3 \pi}{4}$, henceforth in all the plots we set $\theta = 3\pi /4$. Variation of the energy level shift (per unit $\left(\frac{\lambda^2 \omega_0}{16 \pi}\right)$) with respect to acceleration for both configurations is shown in Figure (\ref{eva}).
The resonance energy level shift first increases for lower values of acceleration to reach a peak value, and then decreases with increase in acceleration. The figure shows that the magnitude of energy level shift varies depending on cavity parameter and spatial geometry of a pair of atoms.

\begin{figure}[!htb]
\centering
\includegraphics[scale=0.4]{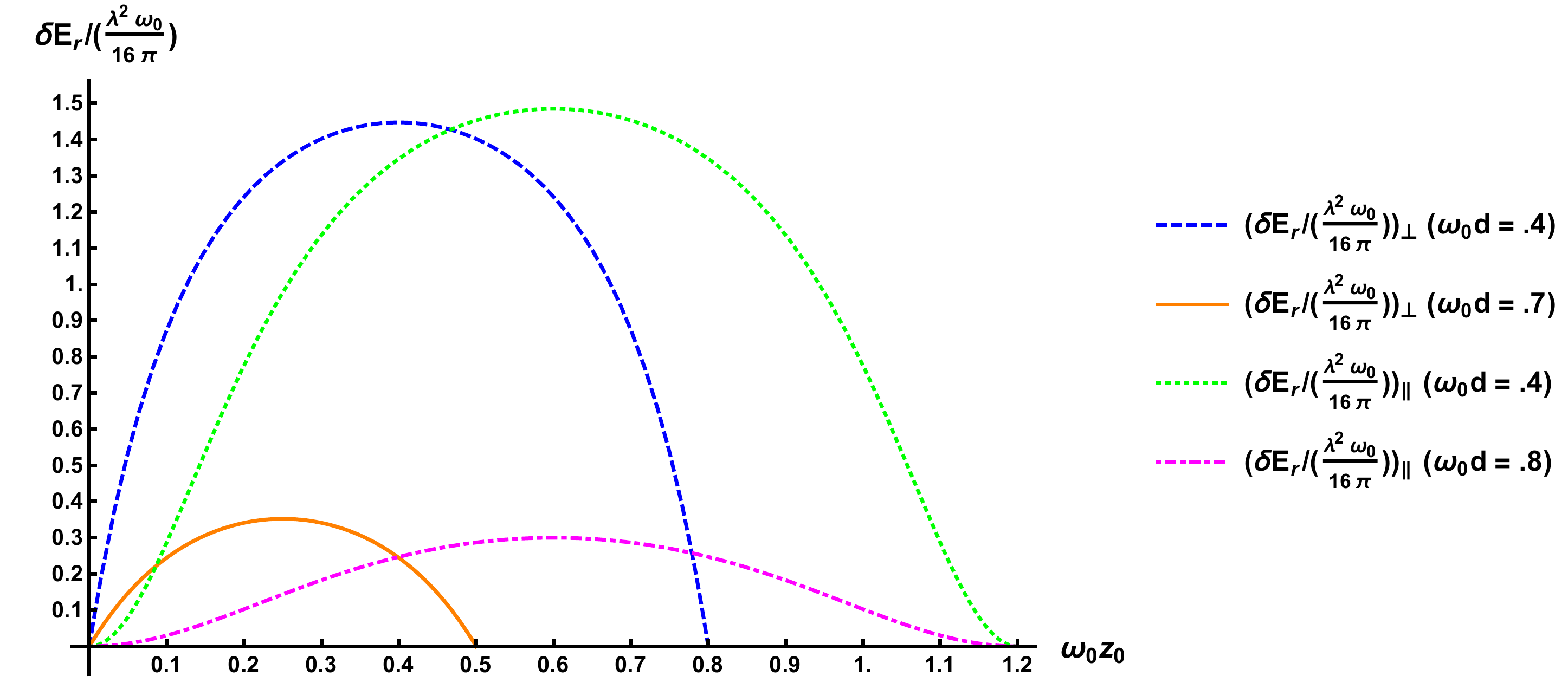}
\caption{(Color Online)  Resonance energy level shift (per unit $\left(\frac{\lambda^2 \omega_0}{16 \pi}\right)$) versus distance of any  one atom from one plate (both configurations), $\theta = 3 \pi / 4$, $a/\omega_0 = 4$, $\omega_0 L = 1.2$.}
\label{evz}
\end{figure}

\begin{figure}[!htb]
\centering
\includegraphics[scale=0.4]{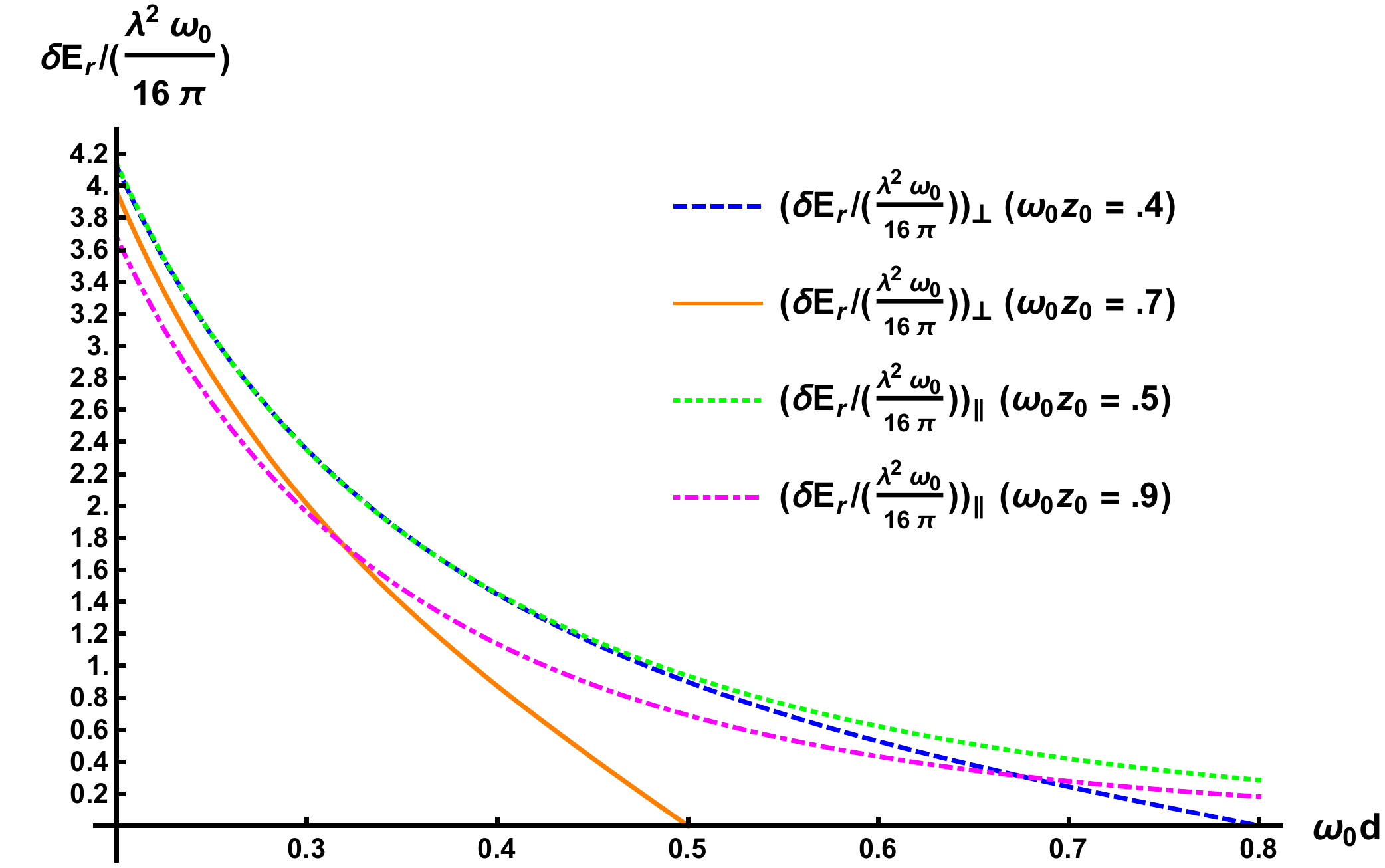}
\caption{(Color Online)  Resonance energy level shift (per unit $\left(\frac{\lambda^2 \omega_0}{16 \pi}\right)$) versus interatomic  distance (for both configurations),  $\theta = 3 \pi / 4$, $a/\omega_0 = 4$, $\omega_0 L = 1.2$. }
\label{evd}
\end{figure}

Let us now study how resonance energy level shift is affected by interatomic distance and their position with respect to the plates. Figure (\ref{evz}) show the variation of energy level shift (per unit $\left(\frac{\lambda^2 \omega_0}{16 \pi}\right)$) with respect to the distance of one atom from the adjacent plate for \textit{configurations 1} and \textit{2}, respectively, for different values of interatomic separation.
From the plots we see  that both the energy shifts and their difference due to different interatomic separation get enhanced when atoms are farther from the bounderies, and diminish as they get closer to the boundary. The energy shift becomes maximum when the atoms are equidistant from both plates (true for both configurations). If either of the atoms touches the plate, resonance interaction will vanish (as can be seen from the relevant mathematical expressions, as well). Figure (\ref{evd}) shows the variation of resonance energy level shift (per unit $\left(\frac{\lambda^2 \omega_0}{16 \pi}\right)$) with respect to interatomic distance for \textit{Configurations 1} and \textit{2} respectively. The energy shift decreases monotonically with increase in interatomic distance for both configuration, though  more sharply in case of \textit{Configuration 1}. This is due to the fact that the atoms move closer to the boundary with increase in $d$ in case of \textit{Configuration 1}, whereas, distance from the boundary remains constant with increase of $d$ in case of \textit{Configuration 2}.

As we have seen, the resonance energy shift  contains boundary dependent terms (Eqs. \ref{shftn},\ref{shftp}).
We now obtain the limiting cases of these expressions to describe the single mirror and free space scenarios.
 $L \rightarrow \infty$ reduces our system to the case of a two-atom-field system with a single mirror boundary.  Taking
the limit  $L \rightarrow \infty$ in Eqs. (\ref{shftn},\ref{shftp}) we obtain 
\begin{equation}\label{sin1}
 \begin{split}
(\delta E_r)_{\perp} = - \dfrac{\lambda^2 \sin 2\theta}{16 \pi} \Big[\dfrac{\cos (\dfrac{2\omega_0}{a}\sinh^{-1}(\dfrac{d a}{2}))}{d \sqrt{1+\dfrac{d^2 a^2}{4}}}  -  \dfrac{\cos (\dfrac{2\omega_0}{a}\sinh^{-1}(\dfrac{D_1 a}{2}))}{D_1 \sqrt{1+\dfrac{D_1^2 a^2}{4}}} \Big]
\end{split}
 \end{equation}
  where $D_1 =  d+2z_0$, and the corresponding $(\delta E_r)_{\parallel}$ is obtained by replacing $D_1$ \textbf{in eq. }(\ref{sin1}) by $D_2 = \sqrt{d^2 + 4 z_0^2}$. These two expressions match with the corresponding expressions obtained in an earlier work \cite{pas3} where the resonance energy shift of a correlated two-atom system was studied in presence of a single mirror. Next, considering the limits $L \rightarrow \infty$ and $z_0 \rightarrow \infty$ together 
reduces our system to a two-atom-field system accelerating in free space. Taking these limits in Eqs. (\ref{shftn},\ref{shftp}), 
yields 
 \begin{equation}\label{free1}
 \delta E_r = - \dfrac{\lambda^2 \sin 2\theta}{16 \pi} \cdot \dfrac{\cos (\dfrac{2\omega_0}{a}\sinh^{-1}(\dfrac{d a}{2}))}{d \sqrt{1+\dfrac{d^2 a^2}{4}}}
 \end{equation}
for both the \textit{Configurations 1} and \textit{2}. Here again we recover the result of earlier works \cite{pas2,zhou} which considered
the resonance energy shift of a system of two accelerating atoms in free space.

\subsection{Relaxation rate of energy due to resonance interaction}

Applying Eq. (\ref{exch}) for the case of the two particles A and B, we obtain the relaxation rate of energy $R_r$ for the two atom system, given by 
 \begin{equation}
\begin{split}
R_r = -i\lambda^2 \int^{\tau}_{\tau_{0}} d\tau' [\chi^{F}(X_{A}(\tau),X_{B}(\tau')) \dfrac{d}{d\tau} C_{A,B}(\tau,\tau') + A\leftrightarrows B \  \text{term}]. 
\end{split}
\end{equation}
Taking the limits $\tau \rightarrow \infty$ , $\tau_0 \rightarrow -\infty$,  and using Eqs.(\ref{chno},\ref{chpa},\ref{cab}), we 
obtain
\paragraph*{\textbf{Configuration 1:}} \begin{equation}\label{raten}
\begin{split}
(R_r)_{\perp} = - \dfrac{\lambda^2 \omega_0 \sin 2\theta}{8 \pi} \sum_{n=-\infty}^{\infty} \Big[\dfrac{\sin (\dfrac{2\omega_0}{a}\sinh^{-1}(\dfrac{z_1 a}{2}))}{z_1 \sqrt{1+\dfrac{z_1^2 a^2}{4}}}  -  \dfrac{\sin (\dfrac{2\omega_0}{a}\sinh^{-1}(\dfrac{z_2 a}{2}))}{z_2 \sqrt{1+\dfrac{z_2^2 a^2}{4}}} \Big]
\end{split}
\end{equation}
\paragraph*{\textbf{Configuration 2:}} \begin{equation}\label{ratep}
\begin{split}
(R_r)_{\parallel} = - \dfrac{\lambda^2 \omega_0 \sin 2\theta}{8 \pi} \sum_{n=-\infty}^{\infty} \Big[\dfrac{\sin (\dfrac{2\omega_0}{a}\sinh^{-1}(\dfrac{z_3 a}{2}))}{z_3 \sqrt{1+\dfrac{z_3^2 a^2}{4}}}  -  \dfrac{\sin (\dfrac{2\omega_0}{a}\sinh^{-1}(\dfrac{z_4 a}{2}))}{z_4 \sqrt{1+\dfrac{z_4^2 a^2}{4}}} \Big].
\end{split}
\end{equation}

The rate of change of energy due to resonance interaction contains boundary dependent terms. Taking the limit
 $L \rightarrow \infty$, we obtain the expressions for the rate corresponding to  an accelerating  two-atom-field system with a single mirror boundary, given by
 \begin{equation}\label{sinr1}
 \begin{split}
(R_r)_{\perp} = - \dfrac{\lambda^2 \omega_0 \sin 2\theta}{8 \pi} \Big[\dfrac{\sin (\dfrac{2\omega_0}{a}\sinh^{-1}(\dfrac{d a}{2}))}{d \sqrt{1+\dfrac{d^2 a^2}{4}}}  -  \dfrac{\sin (\dfrac{2\omega_0}{a}\sinh^{-1}(\dfrac{D_1 a}{2}))}{D_1 \sqrt{1+\dfrac{D_1^2 a^2}{4}}} \Big]
\end{split}
 \end{equation}
 with $D_1 =  d+2z_0$, and the corresponding $(R_r)_{\parallel}$ is obtained by replacing $D_1$ in eq.(\ref{sinr1}) by $D_2 = \sqrt{d^2 + 4 z_0^2}$. On the other hand, taking the limits $L \rightarrow \infty$ and $z_0 \rightarrow \infty$ together leads to the expression for energy relaxation rate in free space \cite{zhou} valid for both configurations, given by 
 \begin{equation}\label{free2}
 R_r = - \dfrac{\lambda^2 \sin 2\theta}{8 \pi} \cdot \dfrac{\sin (\dfrac{2\omega_0}{a}\sinh^{-1}(\dfrac{d a}{2}))}{d \sqrt{1+\dfrac{d^2 a^2}{4}}}
 \end{equation}
 
Similar to the case of energy shift, we study variation of the rate of change of energy due to resonance interaction (for both configurations) with various geometric parameters such as the atom-plate distance ($d$),  and separation between the plates ($L$). We find similar qualitative behaviour as expected from the corresponding expressions of the two physical quantities. However, there exist certain interesting quantitative differences, as noted below.

\begin{figure}[!htb]
\centering
\includegraphics[scale=0.4]{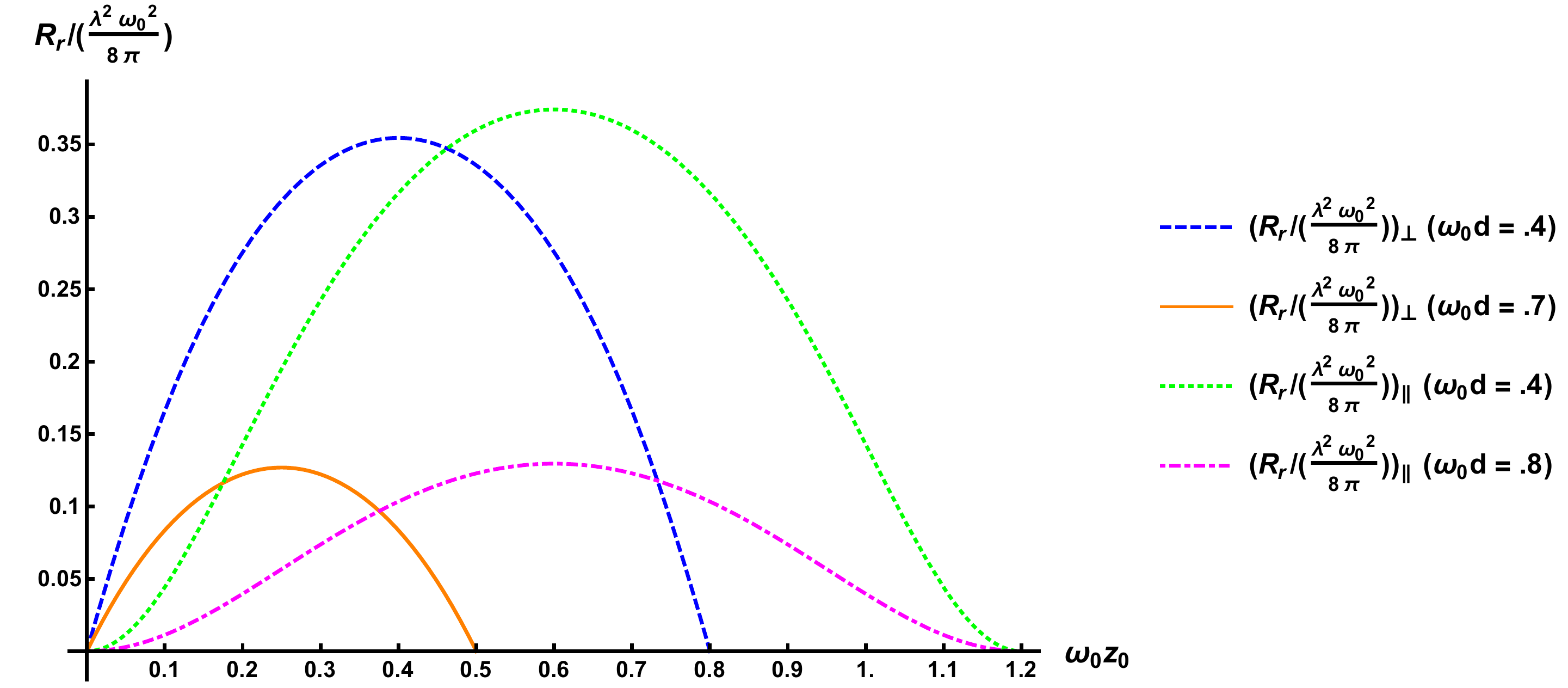}
\caption{(Color Online) Relaxation rate due to resonance interaction (per unit $\left(\frac{\lambda^2 \omega_0^2}{8 \pi}\right)$) versus distance of any one atom from one plate (atoms aligned perpendicular to the plates), $\theta = 3 \pi / 4$, $a/\omega_0 = 4$, $\omega_0 L = 1.2$. }
\label{rvz}
\end{figure} 

 Plots of the energy relaxation rate (per unit $\left(\frac{\lambda^2 \omega_0^2}{8 \pi}\right)$) versus the atom-plate distance for maximal atomic entanglement ($\theta = 3\pi /4$), also show similar features (see Fig. (\ref{rvz}) in  comparison to Fig.(\ref{evz})).

\begin{figure}[!htb]
\centering
\includegraphics[scale=0.4]{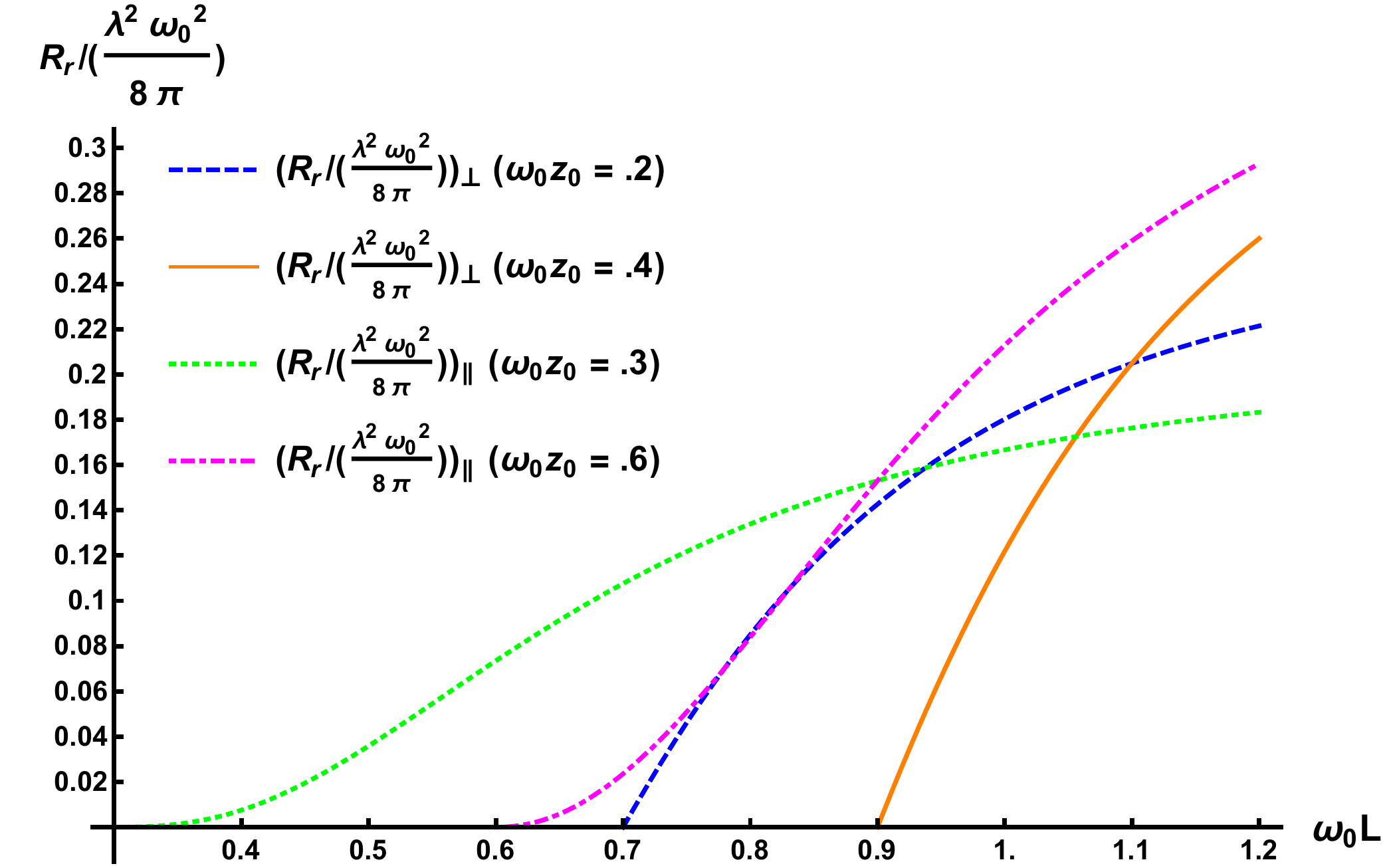}
\caption{(Color Online) Relaxation rate due to resonance interaction (per unit $\left(\frac{\lambda^2 \omega_0^2}{8 \pi}\right)$) versus separation between two mirrors (for both configurations), $\theta = 3 \pi / 4$, $a/\omega_0 = 4$, $\omega_0 d = 0.5$.}
\label{rvl}
\end{figure}

We finally study the change in relaxation rate due to change in the distance between the plates.  Fig. (\ref{rvl})) show the variation of rate of change of energy due to resonance interaction (per unit $\left(\frac{\lambda^2 \omega_0^2}{8 \pi}\right)$) with respect to the separation between plates for \textit{configurations 1} and \textit{2},  respectively. For both configurations the rate increases with increase in separation between plates and tends towards the saturation value of the single boundary limit. This happens as increase in the plate separation makes our system  access a larger number of modes. The variation is steeper for larger atom-plate distance, as seen from the plots.


\section{Conclusions}

The resonance interaction of accelerating atoms with boundary conditions is an interesting and phenomenologically relevant domain of current research. In this work we have studied the resonance interaction between two two-level atoms accelerating between two parallel mirrors. Resonance interaction arises due to coupling of the atoms' internal degrees of freedom with the quantized vacuum. When one atom is in the ground state and the other is in the excited state, they exchange photons between them via the quantized field they are coupled to. 

The boundary conditions imposed by the mirrors modulates the resonance interaction. We have considered two different configurations -- the line joining the two atoms is perpendicular to the axis of the plates in one case, and parallel to the axis of the plates in the other case. We have evaluated the energy level shift and relaxation rate of energy due to resonance interatomic interaction adopting the DDC formalism \cite{dal1,dal2}. Our analysis generalizes earlier calculations of the above quantities  for accelerated two-atom systems near a  single
mirror and in free space \cite{pas2,pas3,zhou}, and  obtains them as  limiting cases.

The goal of the present work has been to study how radiative processes of the
two-atom entangled state undergoing non-inertial motion under boundary conditions can be manipulated by the atomic properties, as well as the atom-plate 
spatial configuration. To this end, we have implemented the appropriate Wightman function for this set-up arising from the individual reflections of photons between the two mirrors. Following the DDC formalism, we have shown that resonance interactions between two atoms can occur even when they are non-maximally entangled. Reassuringly, our result is consistent with what one would expect for a separable state where the interatomic distance dependent energy shift
vanishes at the second order in the matter-field coupling.

We have investigated the variation of the resonance energy shift and relaxation rate with respect to the acceleration,  interatomic separation, distance of the atoms from the mirrors, and separation between the plates. We have shown that resonance interaction can be enhanced or diminished by choosing appropriate values of the initial atomic entanglement, acceleration of atoms, as well as the spatial configuration and geometry of the atom-plate system. During the last decade there is an increasing trend of miniaturizing the cavity dimension in  atom-cavity experiments \cite{vetsch}, \cite{goban} with nanostructured material providing an excellent platform for such efforts.

Before concluding, it may be worthwhile to present a quantitative estimate of our calculated resonant energy shift with presently achieved values of acceleration in current experiments. If the separation of mirrors is chosen of the order of $50 nm$, interatomic distance and distance of the nearest atom and the mirror are chosen in the order of $20 nm$ and $12 nm$, respectively, and the acceleration is chosen in the order of $10^{17} m/s^2$ (achieved by parametric amplifier in superconducting circuits\cite{hac}),  the correction in resonance energy shift due to acceleration will be of the order $10^{-11}$ eV (obtained using equation (\ref{lim}) with $\omega_0 = 5$ eV and $\lambda = 0.1$). Here it may be noted  that Lamb shift of the $\sim 10^{-6} \ eV$ has been measured experimentally \cite{lamb}. It therefore should not be impossible to conceive of future experiments  to measure and verify the results predicted by our analysis.

 We conclude by reemphasizing that our analysis sets up a platform to study if resonance
interaction of accelerating quantum emitters coupled to the electromagnetic vacuum could be controlled in practical devices such as  waveguides. It may be noted that in the present paper our aim has been to explore interesting resonance phenomena under the combined effect of acceleration and boundary conditions. In a future extension of our analysis the effects of environment induced decoherence may be studied in the context of realistic experimental setups. It would be worthwhile to apply the present formalism  \cite{dal1,dal2}  for  radiative emission of atoms  under various
boundary conditions in realistic contexts of quantum transport \cite{trans1}, dynamical Casimir effect, and in case of black hole or cosmological spacetime with generic static or dynamical boundary. \par

\textit{Acknowledgements}: ASM acknowledges support from the project \\ DST/ICPS/QuEST/Q98 from the Department of Science and Technology, India.

\end{document}